\documentclass[conference]{IEEEtran}
\usepackage{times,amsmath,amsfonts,amssymb,amsthm}
\usepackage{cite}
\usepackage{graphicx,subfigure,psfrag}
\usepackage{algorithm,algorithmic}
\usepackage{booktabs}
\usepackage{multirow}
\usepackage{multicol}
\usepackage{colortbl}
\usepackage{color}
\def\lambdabf{\boldsymbol \lambda }

\def\abf{{\bf a}}
\def\bbf{{\bf b}}
\def\cbf{{\bf c}}
\def\dbf{{\bf d}}

\def\gbf{{\bf g}}

\def\Sbf{{\bf S}}

\def\Ac{{\cal A}}

\def\Dc{{\cal D}}

\def\Kc{{\cal K}}

\def\Nc{{\cal N}}

\def\Sc{{\cal S}}

\def\Wc{{\cal W}}

\def\ie{{\it i.e.,\ \/}}
\def\nn{\nonumber}

\newcommand{\Rmnum}[1]{\expandafter\@slowromancap\romannumeral #1@}

\newtheorem{lemma}{Lemma}
\newtheorem{theorem}{Theorem}

\begin{document}

\title{Optimal Cache Placement  for Modified Coded Caching with Arbitrary Cache Size}

\author{\IEEEauthorblockN{Yong Deng
and Min~Dong}
\IEEEauthorblockA{Dept. of Electrical, Computer and Software Engineering, University of Ontario Institute of Technology, Ontario, Canada}
%\IEEEauthorblockA{Email: \{yong.deng, min.dong\}@uoit.ca}
}

\maketitle

\allowdisplaybreaks

\begin{abstract}
We consider content caching between a service provider and multiple cache-enabled users, using the  recently proposed modified coded caching scheme (MCCS) that provides an improved delivery strategy for random user requests. We  develop the optimal cache placement solution for the MCCS  with arbitrary cache size by formulating the cache placement as an optimization problem to minimize the average rate during the delivery phase under random user requests. Through reformulation, we show that the problem is a linear programming problem. By exploring the properties in the caching constraints, we obtain the optimal cache placement solution in closed-form. We verify that the existing cache placement scheme obtained at specific cache sizes is a special case of our solution.
Numerical studies show  how the caching gain changes as the user population increases,  as a result of different cache placement patterns.

  \end{abstract}
%
%%
%% Note that keywords are not normally used for peerreview papers.
%\begin{IEEEkeywords}
%Vehicular Cloud Computing, Tendering, Incentive Mechanism, Privacy Preservation,
%\end{IEEEkeywords}

\section{Introduction}\label{sec:introduction}
%The drastic increase of data-intensive new applications  has caused a shift of wireless traffic  to content-based data for access and sharing.

The last decades have witnessed a dramatic surge of wireless
traffic due to the proliferation of mobile devices \cite{indexglobal}.
Facing the drastic increase of data-intensive new wireless applications and services, future wireless communication networks need to effectively manage  the data traffic congestion  and meet the requirements of timely content delivery. Using network storage resources for content caching has emerged as a compelling technology   to alleviate the network traffic load  and reduce the content access latency for users \cite{Bastug&etal:COMMag14,Wang&etal:COMMag14}. The availability of local caches at the network edge, either at base stations or users, creates new network resources and opportunities to increase the user service capacity. Caching technologies are anticipated to become key technological drivers for  content delivery  in future wireless networks.

%   New Mobile Edge Computing (MEC) has emerged as a promising paradigm
% to handle the challenges arisen\cite{Liang2017:MECbook}.
% At the same time, there is a great pressures confronted by the service providers (SPs) residing among the MEC networks, mainly caused by the large volume of data needed to be moved into and out of the cloud.
% In reality, the capacity of the backhaul that connect the SPs and Edge Nodes (ENs) tends to be limited and has became a bottleneck  to process the explosive wireless data .

%   However, a key observation is that the the users' requests are highly repetitive.
% With this vision,
% placing some cache at ENs tends to be an effective way to alleviate the unprecedent
% backhaul load. It is forecast that the backhaul load can be reduced up to
% $35\%\) by deploying caches at the network edge. Moreover, the user experiences
%will also be improved because the lower latency between ENs and mobile users.
%Obviously, a well designed cache scheme that brings relief to the backhaul %load is essential for a robust MEC system.

The caching design and analysis have attracted increasing
research interests. Many recent works have investigated into the strategies for cache placement and delivery  to understand the effect of caching on reducing the network load \cite{borst2010distributed,Park&Simone:TWC2016,Maddah-Ali&Niesen:TIT2014,Niesen&Maddah-Ali:TIT2015}. Conventional uncoded caching can improve the hit rate  \cite{borst2010distributed,Park&Simone:TWC2016}, but is   not efficient when there are multiple caches\cite{Niesen&Maddah-Ali:TIT2017}.
 Coded caching is first introduced in\cite{Maddah-Ali&Niesen:TIT2014}, where a \emph{Coded Caching Scheme}  (CCS) is proposed  that combines a cache placement  scheme   specifying the cached  (uncoded) content  and a coded multicasting delivery strategy,  assuming uniform file and cache characteristics. With a focus on the theoretical limit, the minimum peak traffic load under caching is analyzed and  substantially coded caching gain for load reduction  is shown. Coded caching has since drawn considerable attentions, with extension to the decentralized cache placement scheme \cite{Niesen&Maddah-Ali:TIT2015}, transmitter caching in mobile edge networks \cite{Sengupta&etal:TIT17,Maddah-Ali&etal:ISIT}, and for both transmitter and receiver caching in wireless interference networks \cite{Xu&etal:TIT17}. Instead of designing cache placement schemes to reduce the peak rate (load), the  optimization of cache placement in the CCS is considered  in \cite{Daniel&Yu:Arxiv2017,Jin&Cui:Arxiv2017},
where using different approaches, the authors have obtained the optimal cache placement strategy to minimize the peak rate.

The  original CCS\cite{Maddah-Ali&Niesen:TIT2014} aims to minimize the  peak rate in the worst-case scenario where users request distinct files, assuming more files  ($N$) than the user population ($K$). For the general cases where multiple users may request the same file,    the CCS contains some redundancy  in the delivery phase causing additional traffic load.
To address this limitation, for   the cache size being multiples of $N/K$, a recent study  \cite{Yu&Maddah-Ali:TIT2018} has proposed a \emph{Modified Coded Caching Scheme} (MCCS)  to  remove the redundancy  of the CCS in the delivery phase, and the minimum average rate  has been obtained. The  corresponding  cache placement has been further shown to be optimal \cite{Jin&Cui:Arxiv2018}.
 However, the technique  used to verify the  optimality of the cache placement     is complicated and is developed only   for the specific cache sizes (multiples of $N/K$), and cannot obtain the general optimal cache placement solution  for the MCCS  with an arbitrary cache size.
%\textcolor{blue}{Note that,  existing studies focus on theoretical understanding of the minimum rate a cached system can achieve. For this purpose, for  cache sizes between multiple of $N/K$, the memory sharing argument is used  to achieve the minimum average rate   \cite{Yu&Maddah-Ali:TIT2018}.  However, memory sharing of different sizes is not a practically valid cache placement scheme, since the cache size is fixed in reality instead of a variable one.}

In this paper, we  use the optimization approach to obtain the optimal  cache placement solution   for the MCCS with any  cache size. We formulate  the cache placement design  into a cache placement optimization problem, aiming to minimize the average rate   in  the delivery phase, under random user demands. Through reformulation,  the optimization problem is shown to be a linear programming problem. By exploring the properties in the problem, we  derive the optimal cache placement solution in closed-form. Our result is general to show the   optimal cache placement solution for the MCCS with arbitrary user population and  cache memory size. We verify that the existing optimal cache placement scheme \cite{Yu&Maddah-Ali:TIT2018,Jin&Cui:Arxiv2018} for  cache sizes at multiples of $N/K$ are special cases in our solution.
%\footnote{The optimal cache placement solution for the CCS has been obtained through optimization in \cite{Daniel&Yu:Arxiv2017}. However, different approach is used there which cannot be applied to solve the cache placement problem for the MCCS.}
% \textcolor{blue}{Note that our optimization approach can also be used to derive the optimal
% cache placement solution that achieves minimum peak rate for both MCCS and CCS.}
Through simulation, we   analyze the performance
of the optimal caching scheme and compare it other schemes. We show how the caching gain changes as the user population increases,  as a result of different cache placement patterns.

%%%%%%%%%%%%%%%%%%%%%%%%%%%%%%%%%%%%%%%%%%%%%%%%%%%
\section{System Model }\label{sec:model}
\begin{figure}[t]
 % \psfrag{Rate}{$\bar R_{\textrm{uni}}$}
  \centering
  \includegraphics[scale=0.35]{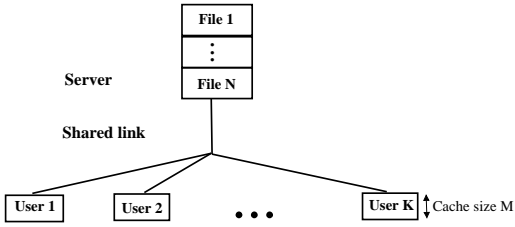}
  \caption{A cache-aided system with end users each equipped with a local cache connecting to the server via a shared link.  } \label{fig:sys_mod}\vspace*{-1.8em}
\end{figure}

Consider a cache-aided transmission system with a server
connecting to $K$ users, each with a local cache, over a shared error-free link, as shown in Fig~\ref{fig:sys_mod}.
%Examples of the described  scenario include  mobile edge computing networks %where network edge nodes (\eg base stations) with cache storage are
%connected through backhaul to a service provider  residing in the cloud, %or  a  server in a base station serving its in-cell users   with local caches.
The server has a database consisting of $N$ files, $\{W_1,\ldots,W_N\}$, each of size $F$ bits.  Denote  $\Nc\triangleq\{1, \ldots, N\}$. We assume  uniform popularity distribution
of these files with $p_n = 1/N$, for $n\in \Nc$.
 Denote the set of users by $\Kc\triangleq\{1,\ldots,K\}$. Each user $k$  has a local cache of  capacity $MF$ bits, for $M \in [0, N]$, and we denote its  cache size (normalized by file size) by $M$.

% simplified MEC network with one SP resides in the cloud and $K$ ENs
% connected to the SP through a shared, error-free backhaul link, as shown in Fig. 1. Each EN serves a number of  users based on their location proximity. Upon
% users' file requests, each EN requests  the files from the SP which contains a copy of the whole library. To mitigate the load of backhaul traffic,
% the ENs may pre-load some content caches during idle time. The SP delivers a coded message that enables all the ENs to reconstruct the files they demand by combining their pre-fetched caches and the coded message.

% \begin{figure}
%  % \psfrag{Rate}{$\bar R_{\textrm{uni}}$}
%   \centering
%   \includegraphics[scale=0.3]{figure1.eps}
%   \renewcommand{\figurename}{Fig.}
%   \caption{An example of cache-aided systems, where edge nodes are connected
% to the central service provider   through a backhaul link. Each edge node has a local cache to alleviate the burden of the backhaul.} \label{fig:sys_mod}
% \end{figure}

%\subsection{System model}
%
%We denote $W_{n}$ as file $n\subseteq\Nc$.  same cache capability $Z_{k}=MF$ bits

The system operates in two phases: cache placement phase
and content delivery phase. The cache placement is performed in advance during the off-peak hours without knowing the user file requests, and is changed at a longer time scale. During this phase, under a  cache placement scheme, each user $k$ uses a caching function $\phi_{k}(\cdot)$
to map $N$ files into its cached content: $Z_{k}\triangleq\phi_{k}(W_1,\ldots,W_N)$.
Each user $k$ independently requests one file  from the server, with the index of requested file denoted by $d_k$, $k\in \Kc$. Denote $\dbf\triangleq[d_1,\ldots,d_K]$ as  the demand vector containing the indices of file requested by all users.
In the content delivery phase, based on the demand vector $\dbf$ and the cache placement, the server generates coded messages and transmit them  to the
users over the shared link.  Denote the codeword as
$X_{\dbf}=\psi_{\dbf}(W_1,\ldots,W_N)$, where $\psi_{\dbf}(\cdot)$ is the encoding function for demand $\dbf$.
Upon receiving the codewords, each user $k$ applies a decoding function $\varphi_{\dbf,k}(\cdot)$ to obtain the (estimated) requested file $\hat{W}_{\dbf,k}$ from the received signal and its cached content  as $\hat{W}_{\dbf,k}\triangleq\varphi_{\dbf,k}(X_{\dbf},Z_{k})$. Thus, an entire coded caching scheme can be represented by
the caching, encoding and decoding functions.

%\emph{Coded Caching}:
In both the CCS \cite{Maddah-Ali&Niesen:TIT2014,Niesen&Maddah-Ali:TIT2015} and the MCCS \cite{Yu&Maddah-Ali:TIT2018},  each file $W_n$ is partitioned into non-overlapping subfiles with equal size, one for each specified user subsets. During the cache placement phase, user $k$ caches those subfiles for the user subsets containing user $k$. In the delivery phase,  the server delivers the missing subfiles of the requested file not in a user's local cache,  using a coded multicasting delivery scheme.
 %%%%%%%%%%%%%%%%%%%%%%%%%%%%%%%%%
\section{Problem formulation}
A key design issue in a coded caching scheme is the cache placement. Existing coded caching schemes describe specific ways of file partitioning for the cache placement, when  cache size $M$ is multiple of $N/K$. Instead of this design approach, we formulate the  coded caching  problem as a cache placement optimization problem for a given cache size $M$, to minimize the average rate (load) over the shared link, where  the delivery strategy is specified by  the MCCS.

\subsection{Cache Placement}
To formulate the problem,  each file $W_n$ is partitioned into $2^K$ non-overlapping subfiles, one for each unique user subset $\Sc\subseteq\Kc $. Since we assume that file lengths and popularity and the cache sizes are all uniform,  a symmetric cache placement is adopted by treating all files equally. Thus,
all the files are partitioned in the same way. That is, let $W_{n,\Sc}$  denote the subfile of $W_n$ for user subset $\Sc$.
Its size satisfies $|W_{1,\Sc}| =\cdots=|W_{N,\Sc}|$, for all $\Sc\subseteq\Kc $. In addition, the size of these subfiles only depends on the size of user subset $|\Sc|$ under the symmetric cache placement.

Note that there  are   ${K \choose l}$ different user subsets with the same size $l$, for $l=0,\ldots,K$, where $l=0$ corresponds to the empty set $\emptyset$.
Let $\Sc^{l}_{i}$ denote user subset $i$ of size $l$, \ie $|\Sc^{l}_i|=l$, for $i=1,\ldots,{K \choose l}$.
 Let $\Ac^l\triangleq\{\Sc^{l}_i, i=1,\ldots,{K \choose l}\}$ denote cache subgroup $l$ containing all user subsets of size
$l$, for $l=0,\ldots,K$.
Thus, all user subsets are partitioned into $K+1$ cache subgroups based on the subset size. Accordingly, all subfiles are  partitioned into $K+1$ subgroups: $\Wc^l=\{W_{n,\Sc}: \Sc \in \Ac^l,  n\in \Nc\}$, $l=0,\ldots,K$, where the subfiles in the same group $\Wc^l$ have the same size.

Define the normalized subfiles size $ a_l \triangleq |W_{n,\Sc}|/F$, as a fraction of  the file size $F$, for all  $n\in \Nc$, $\Sc \in \Ac^l$.
Let $\abf=[a_0,\ldots,a_K]^T$ denote the cache placement  vector (common to all files) describing the   size of subfiles to be cached in  each cache subgroup. Note that $a_0$ represents the fraction of a file
that solely exists in the server. In the cache placement phase, user $k$ caches all the subfiles in $\Wc^l$, $l=1,\ldots,K$, that are for user subsets containing user $k$. In other words, user $k$ caches $\{W_{n,\Sc}:  \Sc \in \Ac^l ~\text{and}~k\in \Sc, n\in \Nc\}$, $l=1,\ldots,K$.

For a given caching scheme, each original file should be able to  be reconstructed by combining all its subfiles. For each file, among the partitioned subfiles, there are  ${K \choose l}$ subfiles with size $a_{l}$ (for all user subsets with $\Sc=l$). Thus, we have the file partitioning constraint
\begin{align}
\setlength{\abovedisplayskip}{1pt}
\setlength{\belowdisplayskip}{1pt}
\sum_{l=0}^{K}{K \choose l}a_l=1.\label{Constraint1}
\end{align}

For the local cache at each user, note that among all user subsets of size $l$, there are total ${K-1 \choose l-1}$  different user subsets containing the same user, for $l = 1,\ldots,K$. Since each file is partitioned based on user subsets, it means that for each file, the total number of subfiles a user can possibly cache is $\sum_{l=1}^{K}{K-1 \choose l-1}$; Considering the subfile size $a_l$ for each cache subgroup $l$, this amounts to  $\sum_{l=1}^{K}{K-1 \choose l-1}a_l$ bits that can be cached by the user for each file. Define $\mu\triangleq M/N$ as the normalized cache size. We have the local cache size constraint at each user as
\begin{align}
\sum_{l=1}^{K}{K-1 \choose l-1}a_l\leq \mu.\label{Constraint2}
\end{align}

\subsection{Content Delivery under the MCCS}\label{sec:deliver}
The recently proposed MCCS  \cite{Yu&Maddah-Ali:TIT2018}
provides a new delivery strategy that removes this redundancy existed in the CCS for further rate reduction.
  The  delivery scheme in the CCS is by multicasting  a unique coded message to each user subset $\Sc\in\Ac^{l+1}$, $l=0,\ldots,K-1$, formed by bitwise XOR operation of  subfiles (of the same size $a_l$) as:
%\begin{align}\label{coded_msg}
$\bigoplus_{k \in \Sc} \! W_{d_k,\Sc\backslash\{k\}}$.
%\end{align}
 Each user in subset $\Sc$ can retrieve the subfile of its requested file.
Assuming the worst case of distinct file requests, coded messages for all user subsets are delivered. There are ${K \choose l+1}$ user subsets in $\Ac^{l+1}$,  to which  coded messages of size $a_l$ are delivered.
 The overall peak rate is $\sum_{l=0}^{K-1}{K \choose l+1}a_l$.

 When the file requests are not distinct, the coded delivery in the original CCS contains some redundant subfiles. Let $\widetilde{N}(\dbf)$ denote the distinct requests for demand vector $\dbf$. Based on the MCCS, it forms a leader group that contains exactly $\widetilde{N}(\dbf)$ distinct requests. Denote $\Dc$ the leader group s.t. $|\Dc|=N_e(\dbf)$. Then any group that has intersection with the leader group is called non-redundant group, which is denoted by $\Sc$. We can know that the number of non-redundant group is ${K \choose l+1}-{K-\tilde{n} \choose l+1}$. From the decentralized MCCS describe in \cite{Yu&Maddah-Ali:TIT2018} users will be able to reconstruct the files they requested once the coded messages of all the non-redundant groups are delivered.

\subsection{Cache Placement Optimization for the MCCS}

%\subsubsection{Expected  rate}
Our objective is to minimize the expected rate $\bar R$ by optimizing the cache placement. From delivering strategy described in Section \ref{sec:deliver}, we know that $\bar R$
is equal to the expected size of all the coded messages of the non-redundant
groups. There are ${K \choose l+1}$
user subsets with size $l+1$, and among them,  ${K-\widetilde{N}(\dbf) \choose l+1}$ are redundant subgroups, of which coded messages to them are redundant for users to recover the subfiles for their requested files. Removing  these redundant transmissions, the expected rate is given by
\begin{align} \label{E(R)}
\bar R=\mathbb{E}_{\dbf}\left(\sum_{l=0}^{K-1}\left[{K \choose l+1}-{K-\widetilde{N}(\dbf) \choose l+1}\right]a_{l}\right)
\end{align}
where the expectation is taken with respect to $\dbf$. Following the common practice, we define
${n \choose k}=0$ when $n<0$ or $k>n$.
The cache placement optimization problem is formulated as
\begin{align}
\textrm{\bf P1}: \;\min_{\abf}\;\; & \bar R \nn\\
\textrm{s.t.} \;\; &
\eqref{Constraint1},\eqref{Constraint2} \nn\\
                   &a_{l}\geq0,\ a_l\leq 1,\ \ l \in \Kc\cup\{0\}\label{UniformCon:3}
\end{align}
where constraints (\ref{UniformCon:3}) are the requirements for the subfile size.
%\label{UniformCon:1-2}
\section{The Optimal Cache Placement for the MCCS}

For the uniform file popularity,
%we assume there will have $u$ distinct requests at each requesting phase.
the probability of having $\tilde{n}$ distinct requests is
$%\begin{align}
P_{\text{u}}(\tilde{n})=\Sbf(K,\tilde{n}) {N \choose \tilde{n}}\frac{\tilde{n}!}{N^K}
$, %\end{align}
for $\tilde{n}=1,\ldots,\min\{N,K\}$, where $\Sbf(\cdot,\cdot)$ is the Stirling number of the second \cite{Riordan2012:StirlingBook}. Based on this, we can express  the expected rate $\bar{R}$ in \eqref{E(R)} as
\begin{align}
\bar{R}=\!\sum_{\tilde{n}=1}^{\min\{N,K\}}\! P_{\text{u}}(\tilde{n})\sum_{l=0}^{K-1}\left[\!{K \choose l+1}-{K-\tilde{n} \choose l+1}\right]a_l.
\end{align}
It is clear that $\bar{R}$ is  linear in $a_l$'s. In addition, all the constraints in {\bf P1} are also linear in $a_l$'s. Thus, {\bf P1} is a linear programming problem with respect to $\abf$.
%Although a linear programming problem can be readily solved numerically by existing software, we note that the problem size increases with.
In the following, we solve  {\bf P1} to obtain the optimal cache placement solution. The result is given bellow.
\begin{theorem}\label{thm1}
For any cache size $M\le N$ and $\mu=M/N$,
%in which $\mu$ is the total memory assigned to one file.
 the optimal cache placement  to minimize the expected rate $\bar{R}$ in {\bf P1} is $\abf^*=[0,\ldots,a^*_{l^*},a_{l^*+1}^*,\ldots,0]^T$, where $\mu K-1 \le l^* < \mu K$, and
%\{0,\ldots,K-1\}$ is the one satisfying  $\frac{l_*}{K}<\mu\leq\frac{l_*+1}{K}$,
\begin{align}\label{opt_a}
% \begin{cases}
a_{l^*}^{*}=  \frac{l^*+1-\mu K}{{K \choose l^*}}, \quad
 a_{l_*+1}^{*}= \frac{\mu K-l^*}{{K \choose l^*+1}}.
%             \end{cases}
\end{align}
The minimum expected rate is
\begin{small}
\begin{align}\label{teoSolu}
\bar{R}^*= &\!\sum_{\tilde{n}=1}^{\min\{N,K\}}\!P_{\text{u}}(\tilde{n})\left(\!
\left[\!{K \choose l^*+1}\!-\!{K- \tilde{n}\ \choose l^*+1}\!\right]\!\frac{l^*+1-\mu K}{{K \choose l^*}}\right. \nn\\
&+\left.\left[{K \choose l^*+2}-{K-\tilde{n} \choose l^*+2}\right]\frac{\mu K-l^*}{{K \choose l^*+1}}\right).
\end{align}
\end{small}
\end{theorem}

We will detail the proof of Theorem \ref{thm1}  in Section~\ref{sec:proof}.

\emph{\textbf{Remark}:} Using a different approach,  \cite{Jin&Cui:Arxiv2018} has shown that the cache placement scheme  proposed by \cite{Yu&Maddah-Ali:TIT2018} for the  cache size at  $M\in\{0,\frac{N}{K},\frac{2N}{K},\ldots,N\}
 $ is optimal for minimizing the expected rate.
However, the approach is more complicated, and cannot be generalized to obtain the optimal cache placement for any arbitrary cache size $M$  between those points.
% \footnote{The argument of memory sharing among two cache sizes used in %\cite{Yu&Maddah-Ali:TIT2018} can theoretically show the achievability of %minimum average rate, but does not provide a practical cache placement for %a given cache size.}.
Our result in Theorem~\ref{thm1} provides   the  optimal cache placement  solution in a closed-form for the MCCS, for any given $M$, $K$ and $N$.
  Also, note that our optimization approach can be applied to derive the optimal cache placement solution for the original CCS  for the peak load minimization, which has been obtained in \cite{Daniel&Yu:Arxiv2017}. However, the approach used there cannot be applied to solve the cache placement problem for the MCCS considered in this work.
%We will show  that  the solution in \cite{Jin&Cui:Arxiv2018} is one  special %case of subfile partitioning in our proof of Theorem~\ref{thm1}.

\subsection{Proof of Theorem~\ref{thm1}} \label{sec:proof}

We first reformulate problem {\bf P1}, and then solve it using the KKT conditions \cite{Boydbook}.
%\subsubsection{Problem reformulation}
Define $\gbf\triangleq[g_0,\ldots,g_K]^T$, where $g_l \triangleq{K \choose l+1}-\sum_{\tilde{n}=1}^{\min\{N,K\}}P_\text{u}({\tilde{n}}){K-\tilde{n} \choose l+1}$, $l=0,\ldots,K$,   $\bbf\triangleq[b_0,\ldots,b_K]^ T$, where $b_l\triangleq{K \choose l}$, and  $\cbf\triangleq[c_0,\ldots,c_K]^T$, where $c_l \triangleq {K-1 \choose l-1}$.
Then, we can rewrite \textrm{\bf P1} as
\begin{align}
\text{\bf P2:}\; \min_{\abf}  \;\; &\gbf^T \abf\ \nn\\
\text{s.t.} \;\;
                   &\bbf^T \abf=1 \label{SimpEqualCon:1}\\
                   &\cbf^T \abf\leq \mu\label{SimpUnequalCon:1}\\
                   &\abf\succeq\ 0,\ \abf\preceq 1.\label{SimpUnequalCon:2}
\end{align}To solve \textrm{\bf P2},
define the Lagrange multipliers  $\lambda_1$, $\lambdabf_2=[\lambda_{2,0},\ldots,\lambda_{2,K}]^ T$, $\lambdabf_3=[\lambda_{3,0},\ldots,\lambda_{3,K}]^ T$ for constraints \eqref{SimpUnequalCon:1} and \eqref{SimpUnequalCon:2}, respectively, and $\nu$ for  constraint \eqref{SimpEqualCon:1}. The KKT conditions are given as follows
\begin{align}
&\eqref{SimpEqualCon:1}, \eqref{SimpUnequalCon:1}, \eqref{SimpUnequalCon:2} \nn\\
& \lambda_{1}^{}(\cbf^T \abf-\mu)=0,\label{KKTcon:inequality1.2}\\
%& a_{l}^{}\geq 0,\quad\; l={0,\ldots, K} \label{KKTcon:inequality2.1}\\
& \lambda_{2,l}^{} a_{l}^{}=0,\quad\; l \in \Kc\cup\{0\} \label{KKTcon:inequality2.2}\\
%& a_{l}^{}\leq 1,\quad\; l={0,\ldots, K} \label{KKTcon:inequality3.1}\\
& \lambda_{3,l}^{}(a_{l}^{}-1)=0,\quad\; l \in \Kc\cup\{0\} \label{KKTcon:inequality3.2}\\
& \gbf+\lambda_1\cbf-\lambdabf_2+\lambdabf_3+\nu \bbf={\bf 0} \label{KKTcon:Lagrange} \\
& \lambda_1 \ge 0, \lambdabf_2 \succcurlyeq {\bf 0},\lambdabf_3 \succcurlyeq {\bf 0}.
\end{align}
%where (\ref{KKTcon:Lagrange}) is the Lagrangian condition.

\subsubsection{Optimal file partitioning strategy}
We first introduce Lemmas \ref{lemmFullUtilize}-\ref{lemmaTwoNon} which help reduce the  complexity in finding the solution.
The  corresponding proofs are omitted due to the space limitation.
\begin{lemma} \label{lemmFullUtilize}
At the optimality, inequality \eqref{SimpUnequalCon:1} is attained with equality, \ie
the cache storage $\mu$ is always fully utilized under the optimal cache placement vector $\abf^*$ for {\bf P1}.
\end{lemma}
%To minimize the expected rate $\bar R$, t
%\emph{\textbf{Remark:}} Lemma \ref{lemmFullUtilize} indicates that for any %optimal cache placement $\abf^*$, we have $\cbf^T\abf^{*}= \mu$.

\begin{lemma}\label{lemmZeroOne}
 When $\mu=0$, the optimal  $\abf^*=[1,0,\ldots,0]^ T$ with the minimum expected rate $\bar{R}^*=E(\widetilde{N}(\dbf))$; when $\mu=1$, the optimal $\abf^*=[0,\ldots,0,1]^T$ with
the minimum expected rate $\bar{R}^* = 0$.
\end{lemma}

Lemma~\ref{lemmZeroOne} describes the two extreme cases of having no cache memory ($\mu=0$) and sufficient cache size to hold all $N$ files ($\mu=1$). In the following, we only need to discuss the case when $\mu\in(0,1)$.

By exploring the properties of KKT conditions \eqref{KKTcon:inequality2.2}-\eqref{KKTcon:Lagrange}, we  show below the condition on $\abf$ for  $\lambda_1$ and $\nu$  having feasible solutions. \begin{lemma}\label{lemmaTwoNon}
For $\mu\in(0,1)$, the optimal cache placement vector $\mathbf{a^{*}}$ has at most two non-zero elements.
\end{lemma}
% \IEEEproof
% \endIEEEproof
Lemma~\ref{lemmaTwoNon} implies that the number of non-zero elements of an optimal caching
vector $\abf^*$ can only be one or two (it cannot be zero due to constraint \eqref{SimpEqualCon:1}):

%\begin{enumerate}
% \emph{ i) No nonzero placement parameter}:
% This means $a_l=0$ for all $l\in\{0,\ldots,K\}$.
% It is obviously not possible because
% (\ref{SimpEqualCon:1}) shows that the summation of all the subfiles should
% be equal to the file size.

\emph{Case 1) One non-zero element}: In this case, there exists $a_l\neq0$ for some $l\in \{0,\ldots,K\}$, and $a_j = 0$ for $\forall j\neq l$, $j\in \{0,\ldots,K\}$.  From \eqref{SimpEqualCon:1}, we have $b_la_{l}=1$.
%\begin{align}
%\label{OneNonzeroEqu1}
%\end{align}
Thus, we have
%\begin{align}
$a_l={1}/{b_l}={1}/{{K \choose l}}$.
%\end{align}
To find the cache  size that leads to this solution, note that since there is only one non-zero subfile size, from Lemma~\ref{lemmFullUtilize}, we have $c_la_l=\mu$. Thus, the relation of the normalized cache size and index $l$ is given by $\mu={c_l}/{b_l}={{K-1 \choose l-1}}/{{K \choose l}}={l}/{K}$.

%{\textbf{Optimal solution:}}
Thus, if for some $l \in \{0,\ldots,K\}$ satisfies $\mu={l}/{K}$,
the optimal cache placement is $\abf^*=[0,\ldots,0,a^*_{l^*},0\ldots,0]^{T}$, where $a^*_{l^*}={1}/{{K \choose l^*}}$. For given demand $\dbf$, the corresponding rate
$R$ can be computed based on the redundancy to be removed in the delivery
phase, and we have
\begin{align}
\hspace*{-.5em}R=\begin{cases}
{\frac{K-i}{i+1}}, \quad\quad\quad\quad\quad\quad~\text{for}~  K-\widetilde{N}(\dbf)< i+1  \\
 {\frac{K-i}{i+1}}\left(1-\frac{(K-i-1)\ldots(K-i-\widetilde{N}(\dbf))}{K(K-1)\ldots(K-\widetilde{N}(\dbf)-1)}\right), \; \textrm{otherwise}.
\end{cases}
\end{align}
As a result, the expected rate is
$%\begin{align}\label{proof:avg_R}
\bar{R}=\sum_{\tilde{n}=1}^{\min\{N,K\}}P_\text{u}({\tilde{n}})R.
$%\end{align}

\emph{\textbf{Remark}:} The optimal solution with one non-zero element in $\abf^*$ corresponds to \emph{equal file partitioning}, where all subfiles have equal size. The optimal $\abf^*$ obtained above exactly matches the cache placement scheme proposed in  \cite{Yu&Maddah-Ali:TIT2018} for cache size at points $M\in\{0,\frac{N}{K},\frac{2N}{K},\ldots,N\}$. We see that it is a special case in our general cache placement optimization problem.
%A different approach  in \cite{Jin&Cui:Arxiv2018} with a different approach %that verify the optimality of these solutions.

% replicates exactly the optimal convex envelop in \cite{Yu&Maddah-Ali:TIT2018},
%which are the solutions for $M\in\{0,\frac{N}{K},\frac{2N}{K},\ldots,N\}$.

\emph{Case 2) Two non-zero elements}: In this case, there exist some $i$ and $j$, such that $a_i, a_j \neq0$, and $a_l = 0$, $\forall l \neq i, j$, $l\in\{0,\ldots,K\}$. With only two non-zero variables $a_{i}$ and $a_{j}$, from Lemma 1, we
can rewrite (\ref{KKTcon:inequality1.2}) as
\begin{align}\label{equ:SolveKKT2.1}
c_ia_{i}^{}+c_ja_{j}^{}-\mu=0.
\end{align}
Also from (\ref{SimpEqualCon:1}), we have
\begin{equation}\label{equ:SolveKKT3}
b_i a_{i}+b_j a_{j}=1.
\end{equation}
From (\ref{equ:SolveKKT2.1}) and (\ref{equ:SolveKKT3}), we have  $a_{i}^{}=\frac{b_j\mu-c_j}{b_jc_i-b_ic_j}$ and $a_{j}^{}=\frac{b_i\mu-c_i}{b_ic_j-b_jc_i}$. Since
$a_{i}$ and $a_{j}$ are both non-zero, the two solutions only exists when
\begin{align}
\frac{b_j\mu-c_j}{b_jc_i-b_ic_j} &> 0, \quad
%\end{align}and
%\begin{align}
\frac{b_i\mu-c_i}{b_ic_j-b_jc_i}> 0.\label{inequal:SolveMu}
\end{align}
Assume $0\leq i< j\leq K$. Since ${c_i}/{b_i}={{K-1 \choose i}}/{{K \choose i}}$ is an increasing function of $i$, we have $b_jc_i-b_ic_j<0$. Based on \eqref{inequal:SolveMu}, we have $b_i\mu-c_i> 0$ and $b_j\mu-c_j<0$, which means $i$ and $j$ should satisfy ${c_i}/{b_i}<\mu<{c_j}/{b_j}$.

% Thus, for user normalized cache size $\mu\in({c_i}/{b_i},{c_j}/{b_j})$, %where $0\leq i< j\leq K$, the optimal
% cache placement is $a_{i}^{*}=\frac{d_{j}\mu-c_j}{d_{j}c_i-b_ic_j}$, $a_{j}^{*}=\frac{b_i\mu-c_i}{b_ic_j-d_{j}c_i}$, %and the rest $a^*_l$'s are all $0$.
% The corresponding expected rate $ {\bar R}=g_ia_i^*+g_ja_{j}^*$.

%Note that the exact values of $i$ and $j$ that fall into $[0,K]$ remain %unknown. We need to find the optimal combination of $i$ and $j$ among all %the possibilities.

\begin{lemma}
For  $i<j$ and $i,j\in\{0,\ldots,K\}$ satisfying $\mu\in({c_i}/{b_i},{c_j}/{b_j})$, the expected rate $\bar{R}$ is a decreasing function of $i$ and an increasing function of $j$.\label{lemmIncresing}
\end{lemma}
From Lemma 4, we have the conclusion that
the minimum expected rate $\min_{0\le i<j\le K}\bar{R}$ can only be obtained when $j=i+1$.
Any other relation would result in larger $\bar{R}$.
%either a smaller $i$ or larger $j$ that generates a larger $\tilde R$.
%Then we can deduce the solutions with two nonzero subfiles.
Following this, for $0\le i\le K-1$ satisfying ${c_i}/{b_i}<\mu<{c_{i+1}}/{b_{i+1}}$, we have the optimal $a^*_i$ and $a^*_{i+1}$ as
\begin{align*}
a_{i}^{*}&=\frac{b_{i+1}\mu-c_{i+1}}{b_{i+1}c_i-b_ic_{i+1}}, \quad
a_{i+1}^{*}=\frac{b_i\mu-c_i}{b_ic_{i+1}-b_{i+1}c_i}.
\end{align*}
The corresponding expected rate is $\bar{R}= g_ia_i^*+g_{i+1} a_{i+1}^*$.
%\emph{\textbf{Optimal solution for non-zero placement parameters:}}
Substituting the values of $g_i$, $b_i$ and $c_i$ into the above expressions, let $l^*=i$, we have the following conclusion: For $\mu K-1 < l^* < \mu K$,
the optimal cache placement $\abf^*=[0,\ldots,a^*_{l^*},a_{l^*+1}^*,\ldots,0]^ T$ where $a^*_{l^*}$ and $a_{l^*+1}^*$ are given as in \eqref{opt_a}.
% \begin{align*}
% a_{l^*}^{*}=\frac{l^*+1-\mu K}{{K \choose l^*}}, \quad a_{l^*+1}^{*}=\frac{\mu K-l^*}{{K \choose l^*+1}}.
% \end{align*}

\emph{\textbf{Remark}}: The optimal cache placement indicates that, each file is split into two parts with sizes $(l^*+1-\mu K)$ and $(\mu K-l^*)$. Then each part is further partitioned into subfiles of  equal sizes, with the first  partitioned into ${K \choose l^*}$ subfiles (for cache subgroup $l^*$), and the second partitioned into ${K \choose l^*+1}$ subfiles (for cache subgroup  $l^*+1$). User $k$ will cache these two types of subfiles for all user subsets including $k$.

Given any $\widetilde{N}(\dbf)$, the  rate $R$ depends on the amount of redundancy removed in the delivery phase in following cases:

%\begin{itemize}
%\item \emph{For $K-\widetilde{N}(\dbf)\geq l^*+2$}: There are redundant coded messages
%for both user subsets of size $l^*$ and $l^*+1$.% \begin{small}
%% \begin{align}\label{TwoRedundancy}
%% \hspace*{-.5em}R=&\left[{K \choose l^*+1}-{K-\widetilde{N}(\dbf) \choose l^*+1}\right]\frac{l^*+1-\mu K}{{K \choose l^*}}\nn\\
%% &+\left[{K \choose l^*+2}-{K-\widetilde{N}(\dbf) \choose l^*+2}\right]\frac{\mu K-i}{{K \choose l^*+1}}.
%% \end{align}
%% \end{small}
%\item \emph{For $K-\widetilde{N}(\dbf)= l^*+1$}:
%The redundant coded message can only be found for user subsets of size $l^*$.
%% \begin{small}
%% \begin{align}
%% R=&\left[{K \choose l^*+1}-{K-\widetilde{N}(\dbf) \choose l^*+1}\right]\frac{l^*+1-\mu K}{{K \choose l^*}}\nn\\
%% &+{K \choose l^*+2}\frac{\mu K-l^*}{{K \choose l^*+1}}\label{OneRedundancy}.
%% \end{align}
%% \end{small}
%\item \emph{For $K-\widetilde{N}(\dbf)< l^*+1$}: No redundant message  for any user subsets.% \begin{align}
%% R\!=\!{K \choose l^*+1}\!\frac{l^*+1-\mu K}{{K \choose  l^*}}\!+\!{K \choose l^{*}+2}\!\frac{\mu K- l^*}{{K \choose l^*+1}}.\label{ZeroRedundancy}
%% \end{align}
%\end{itemize}

\begin{itemize}
\item {$K-\widetilde{N}(\dbf)\geq l^*+2$}: There are redundant coded messages
for both user subsets of size $l^*$ and $l^*+1$, and we have
{\small
\begin{align}\label{TwoRedundancy}
\hspace*{-.5em}R=&\left[{K \choose l^*+1}-{K-\widetilde{N}(\dbf) \choose l^*+1}\right]\frac{l^*+1-\mu K}{{K \choose l^*}}\nn\\
&+\left[{K \choose l^*+2}-{K-\widetilde{N}(\dbf) \choose l^*+2}\right]\frac{\mu K-l^*}{{K \choose l^*+1}}.
\end{align}
}
\item {$K-\widetilde{N}(\dbf)= l^*+1$}:
The redundant coded message can only be found for user subsets of size $l^*$, and we have
{\small
\begin{align}
\hspace*{-.5em}R\!=\!&\!\left[\!{K \choose l^*\!+\!1}\!-\!{K\!-\!\widetilde{N}(\dbf) \choose l^*+1}\!\right]\!\frac{l^*+1-\mu K}{{K \choose l^*}}\nn\\
&+{K \choose l^*+2}\frac{\mu K-l^*}{{K \choose l^*+1}}\label{OneRedundancy}.
\end{align}
}
\item
\emph{$K-\widetilde{N}(\dbf)< l^*+1$}: No redundant message  for any user subsets, and the rate is
{\small
\begin{align}
R\!=\!{K \choose l^*+1}\frac{l^*+1\!-\!\mu K}{{K \choose  l^*}}\!+\!{K \choose l^{*}+2}\frac{\mu K- l^*}{{K \choose l^*+1}}.\label{ZeroRedundancy}
\end{align}
}
\end{itemize}
The minimum expected rate is given by $\bar{R}=\sum_{\tilde{n}=1}^{\min\{N,K\}}P_\text{u}({\tilde{n}})R$. From  \eqref{TwoRedundancy}-\eqref{ZeroRedundancy}, we arrive at the expression in \eqref{teoSolu}.

Cases 1 and 2 give  the optimal cache placement for any $\mu\in(0,1)$.
%Case 1 gives the optimal placement solution for $\mu=\frac{ l^*}{K}$, $l^*\in\{1,\cdots,K-1\}$, %and Case 2 gives the optimal solution for $\frac{l^*}{K}<\mu<\frac{l^*+1}{K}$, % $l^*\in\{0,\ldots,K-1\}$.
 Combining these with the solutions for $\mu=0$ and $\mu=1$ in Lemma \ref{lemmZeroOne}, we have the results in Theorem~\ref{thm1}.

\section{Numerical Results}
\label{sec:performance}

%\setlength{\belowdisplayskip}{0pt}
%We present the numerical results to evaluate the performance of the optimized cache placement scheme.
Consider a system with $N$ files of equal size,
$K$ users with the same cache size $M$.
First, we show in Table \ref{solutionM} the values of the optimal cache placement vector $\abf^*$  for different cache sizes $M=0,\ldots,10$, for $K=7$ and $N=10$.  Beside the two extreme cases of $M=0$ or $10$ (all the
files are  either in the servers, or stored at the local cache),  for $M$ in between, we
 see that $\abf^*$ always has two non-zero
elements (\ie two different subfile sizes for two cache subgroups), and  their locations shift to the  cache subgroup of larger size $l$ as $M$ increases.

% Fig. \ref{figMemoryRate} shows the trade-off of expected rate $\bar{R}$ vs.  cache size $M$ for $N=5$ and $K=10$.
% We compare the performances of the optimal cache placement scheme for the MCCS obtained in Theorem~\ref{thm1}
% and the state-of-art schemes, including the uncoded cache scheme,
% the centralized CCS\cite{Maddah-Ali&Niesen:TIT2014,Daniel&Yu:Arxiv2017},
% the decentralized CCS\cite{Niesen&Maddah-Ali:TIT2015} and the decentralized MCCS\cite{Yu&Maddah-Ali:TIT2018}. As expected,  the  optimal placement solution for the
% MCCS  always outperforms all the other schemes regardless the
% cache size $M$.
% %Specifically, it decreases piece-wise linearly for $M\in [iN/K, (i+1)N/K]$, $i=0,\ldots,K-1$.
% Also, at all the  points $M=iN/K$, $\abf^*$ has one non-zero
% element (\ie equal file partitioning), and for $M$ in between, $\abf^*$ has two non-zero elements (\ie two different subfile sizes for two cache subgroups
% that can also be interpreted as the memory sharing scheme between two different
%$M$'s).

\begin{figure}[h]
% \setlength{\abovecaptionskip}{0pt}
% %\setlength{\belowcaptionskip}{2pt}
% %  \psfrag{Rate}{\hspace*{-2em}Expected rate $\bar R$}
%  % \psfrag{M}{$M$}
%   \centering
%   \includegraphics[scale=.42]{figure2.eps}
%   \caption{ The expected rate $\bar R$ versus $M$  ($K=10$ and $N=5$).}\label{figMemoryRate}
%\end{figure}
%\begin{figure}
%  \psfrag{Rate}{\hspace*{-3em}Expected rate $\bar R$}
%  \psfrag{K}{$K$}
  \centering
  \includegraphics[scale=0.5]{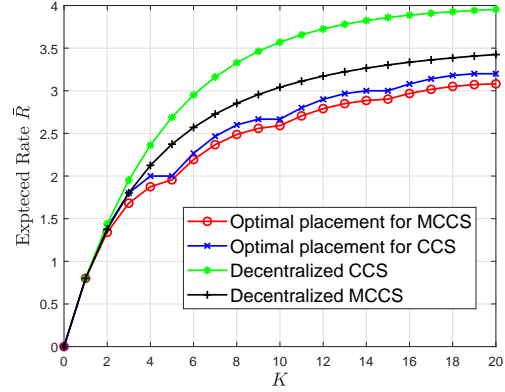}
  \caption{The expected rate $\bar R$ versus $K$ ($M=2$ and $N=10$).}\label{figENRate}\vspace{-1em}
\end{figure}

In Fig. \ref{figENRate}, for fixed  $N=10$ and   $M=2$, we  study how the expected rate $\bar{R}$ changes with the increasing number of users $K$. We compare the optimal cache placement  for the MCCS, with other schemes, including the centralized CCS\cite{Maddah-Ali&Niesen:TIT2014},
and the decentralized CCS\cite{Niesen&Maddah-Ali:TIT2015} and MCCS\cite{Yu&Maddah-Ali:TIT2018}. The optimal cache placement solution achieves the minimum expected rate, and thus  outperforms all the other schemes. For both the CCS and the MCCS, the expected rate under the optimal solution increases with $K$ with a certain pattern. The rate increment slows down when $K$ reaches $i N/M$, for $i=1,2,\ldots$, but becomes higher after $K$ passes those points. This suggests the caching gain increases with $K$  from  $(i-1) N/M$ to $i N/M$, with the highest gain achieved at $i N/M$. Note that
for a given normalized cache size $\mu$, $K$ determines the caching subgroup sizes and the number of user subsets for coded multicasting under the optimal cache placement.
For $K=i N/M$,
the optimal $\abf^*$ only has one cache subgroup (\ie equal file partitioning), while at the two sides of this point,   two different caching subgroups are used.

% For example, when $K=10$, the caching gain is obtained by using the caching subgroup of size $10$. However, when $K=9$, the caching gain obtain by two caching subgroups of sizes $9$ and $10$, and the two caching subgroups for $K=11$ are of size $10$ and $11$.  Since the subfiles  for  cache subgroup size $9$ is small when $K=9$, the rate difference is negligible when $K$ is increased to $10$. The noticeable increase at  $K=11$ is because the caching gain offer by group with size $10$ and $11$ has significant difference compared with the gain offered only by caching subgroup of size $10$.
 \renewcommand{\arraystretch}{0.8}%{1.05}
\begin{table}[t]
\centering\caption{The Optimal cache placement ($K=7$,
$N=10$).}\label{solutionM}
\begin{tabular}{c|c|c|c|c|c|c|c|c|c}
\hline
\multirow{2}{*}{$M$}&\multicolumn{8}{c|}{Optimal Cache Placement}\\ \cline{2-9}  &$a_0$&$a_1$&$a_2$&$a_3$&$a_4$&$a_5$&$a_6$&$a_7$\\ \hline\hline
 0& 1.0& 0&0  & 0& 0 &  0 & 0  &0\\
 1& 0.3& 0.1&0  & 0& 0&0 & 0  &0&\\
 2& 0&0.086 & 0.019 &0  & 0 &0 & 0  &0&\\
 3& 0& 0& 0.043  & 0.003 &  0  &0 & 0 &0&\\
 4& 0& 0&0.01  & 0.023  & 0  &  0& 0 &0&\\
 5& 0& 0&0  & 0.014  &0.014  &  0  &  0  &0&\\
 6& 0& 0&0  & 0  & 0.01  & 0.023&  0  & 0  &\\
 7& 0& 0&0  & 0  &0.003 & 0.043&0 &  0&\\
 8& 0& 0&0  & 0  & 0& 0.019 &0.086  &  0  &\\
 9& 0 & 0&0  & 0  & 0&0 & 0.1&0.3&\\
 10& 0 & 0&0  & 0  & 0&0 & 0  &1.0&\\
\hline\\[-1.9em]
\end{tabular}
\end{table}

\section{Conclusion}\label{sec:conclusion}
In this paper,
we formulate the  general cache placement design  for the MCCS under uniform file popularity as a cache placement optimization problem to minimize the expected rate during the  delivery phase, for any number of users,  files, and cache size. Through the optimization approach and by exploring the property of the optimization problem, we obtained the general optimal cache placement solution in closed-form for the MCCS.
\bibliographystyle{IEEEtran}

\end{document}